\documentclass[11pt]{article}

\usepackage[margin=1in]{geometry}
\usepackage{graphicx}
\usepackage{amsmath, amssymb}
\usepackage{booktabs}
\usepackage{siunitx}
\usepackage{caption}
\usepackage{subcaption}
\usepackage{cite}
\usepackage[hidelinks]{hyperref}
\usepackage{placeins}
\usepackage{needspace}
\usepackage{float}
\title{\textbf{Scaling Limits of Multichannel Spectral Routers for Snapshot Imaging}}
\author{
Junseo Han$^{1,\dagger}$, Seunghyun Lee$^{2,\dagger}$,
Donghyun Kim$^{2}$, and Haejun Chung$^{1,2,*}$\\[0.5em]
{\small $^{1}$Department of Artificial Intelligence Semiconductor Engineering,}\\[-0.15em]
{\small Hanyang University, Seoul 04763, Republic of Korea}\\[-0.05em]
{\small $^{2}$Department of Electronic Engineering,}\\[-0.15em]
{\small Hanyang University, Seoul 04763, Republic of Korea}\\[-0.05em]
{\small $^{\dagger}$These authors contributed equally to this work.}\\
{\small $^{*}$Corresponding author: \texttt{haejun@hanyang.ac.kr}}
}
\date{}

\begin{document}

\maketitle

\begin{abstract}
Inverse-designed spectral routers can enable compact snapshot spectral imaging by directing different wavelengths to designated detector sub-pixels without mechanical scanning or absorptive filters. Here, we examine how routing performance varies as the number of spectral channels increases while the visible bandwidth remains fixed. A delay-bandwidth analysis relates channel count, worst-channel efficiency, and device thickness, explaining the growing optical capacity required to generate many spatially distinct spectral outputs. We then use adjoint-based topology optimization and three-dimensional finite-difference time-domain simulations to design TiO\(_2\)/SiO\(_2\) routers with 9, 16, 25, and 36 channels under matched material, bandwidth, and thickness conditions. The average routing efficiency increases with sub-pixel size and approaches a plateau, while the maximum near-plateau efficiency decreases monotonically from 97.0\% for 9 channels to 82.3\% for 36 channels. At a fixed channel count, compressing the wavelength spacing to 8 nm or rearranging the wavelength-to-sub-pixel assignment changes the efficiency by less than one percentage point. These results indicate that the observed efficiency penalty is governed mainly by the number of wavelength-dependent routing constraints rather than by wavelength spacing or local wavelength arrangement. The findings quantify the trade-off between spectral sampling and optical throughput in compact snapshot spectral imagers.
\end{abstract}

\section{Introduction}

Natural scenes are commonly modeled as radiance fields defined over continuous
spatial and spectral coordinates, whereas practical image sensors reduce this
continuum to a discrete array of spatially integrated and spectrally weighted
measurements~\cite{field1987relations,ruderman1993statistics,
simoncelli2001natural,nascimento2002statistics,shannon1949communication}.
In the spatial dimension, a finite-aperture imaging system imposes a
diffraction- and aberration-dependent optical transfer function with a finite
optical bandwidth, and the resulting irradiance distribution is further
integrated and sampled by finite-area detector
pixels~\cite{hopkins1955frequency,stokseth1969properties,
goodman2005introduction,fossum2016quanta,mahato2018measuring}.
In the spectral dimension, conventional single-chip cameras map continuously
varying scene radiance onto a small number of broad and overlapping detector
responses, typically implemented using red, green, and blue color-filter arrays,
to reproduce color appearance rather than to recover wavelength-resolved
spectra~\cite{bayer1976color,gunturk2005demosaicking,lukac2005color,zou2022pixel}.
As imaging systems become more compact and pixels continue to shrink, the
benefit of finer spatial sampling is increasingly offset by reduced photon
collection, diffraction-limited blur, stronger angular dependence, and optical
crosstalk between neighboring pixels~\cite{lee2024inverse,fossum2016quanta,hirigoyen2008fdtd,vaillant2011characterization,nishiwaki2013efficient,zou2022pixel}.

These trade-offs are particularly important in mobile and pixel-scale imagers,
where the optical stack above each detector pixel must collect, spectrally
select, and confine light within an increasingly small aperture~\cite{fossum2016quanta,hirigoyen2008fdtd,vaillant2011characterization}.
As the pixel size approaches the wavelength scale, the microlens, color-filter,
and pixel-isolation stack must simultaneously maintain high collection
efficiency, preserve angular tolerance, and suppress optical crosstalk, while
the achievable spatial resolution remains bounded by the diffraction-limited
passband of the optical front end~\cite{goodman2005introduction,fossum2005sub,fossum2016quanta,hirigoyen2008fdtd,vaillant2011characterization,kim2025cmos}.
At the same time, absorptive color-filter arrays reject or attenuate a
substantial fraction of out-of-band photons, reducing the detected photoelectron
count and thereby degrading the shot-noise-limited signal-to-noise
ratio~\cite{gunturk2005demosaicking,zou2022pixel,fossum2016quanta,gnanasambandam2022exposure}.
These limitations have motivated sensor-integrated wavelength-sorting optics
that redirect spectral components to designated detector regions rather than
absorbing or rejecting them. Examples include color splitters, color routers,
and metasurface-based wavelength sorters, which can improve photon utilization
relative to conventional absorptive filters~\cite{kim2024freeform,jeon2026inverse,lee2024inverse,
nishiwaki2013efficient,camayd2020multifunctional,miyata2021full,
zhao2021perfect,zou2022pixel,li2022single,catrysse2022subwavelength,
catrysse2023spectral,ding2017beam,lee2019all}.

\begin{figure}[t]
\centering
\includegraphics[width=\linewidth]{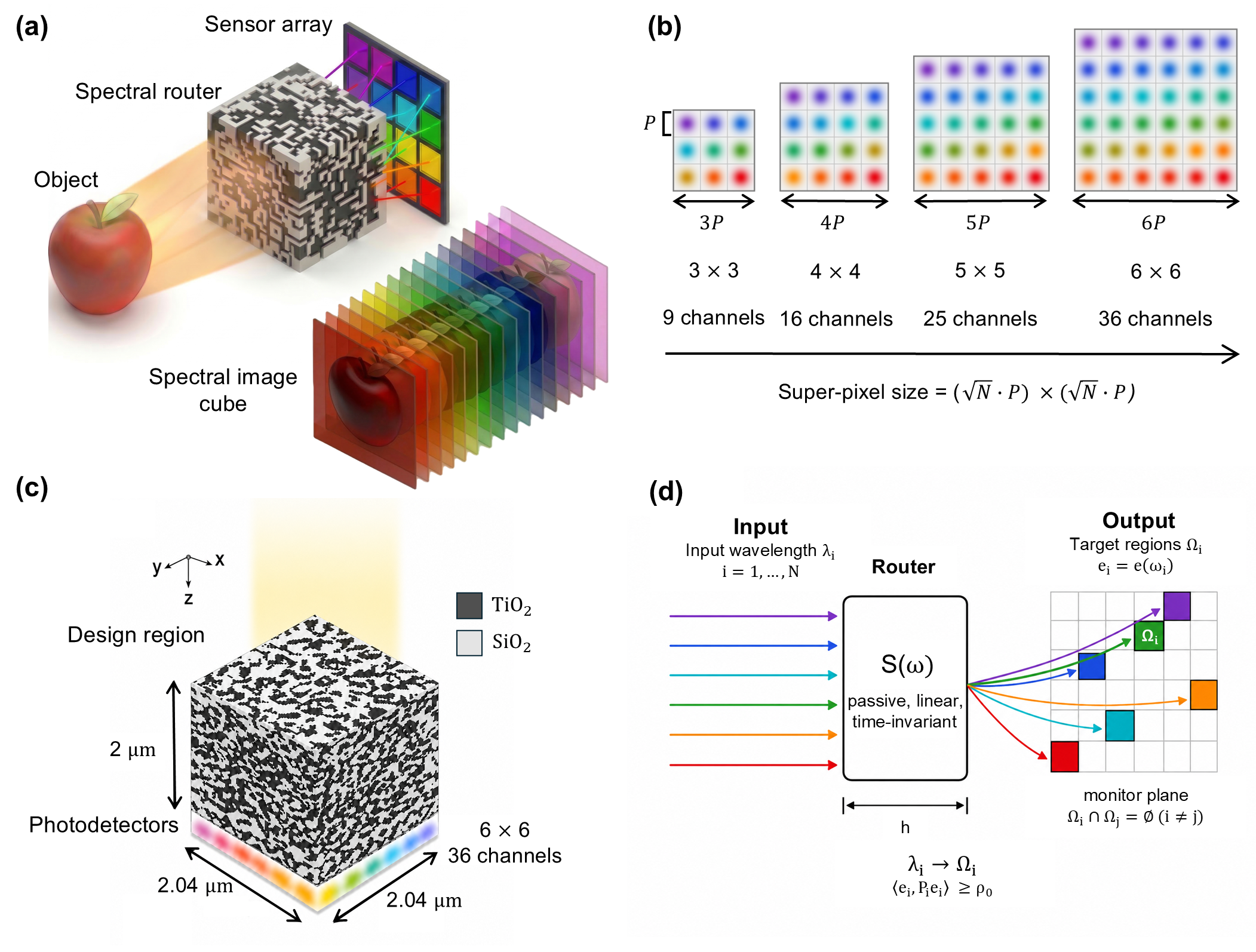}
\caption{
Concept and configuration of the inverse-designed spectral router for snapshot spectral imaging.
(a) Broadband light from an object is directed to wavelength-dependent positions on the sensor array, enabling single-shot acquisition of a spectral image cube.
(b) Square super-pixel configurations with \(N=9\), 16, 25, and 36 channels, arranged as \(3\times3\), \(4\times4\), \(5\times5\), and \(6\times6\) sub-pixel arrays, respectively. Each sub-pixel has side length \(P\), giving a super-pixel size of \((\sqrt{N}P)\times(\sqrt{N}P)\).
(c) Representative binary TiO\(_2\)/SiO\(_2\) structure for the 36-channel router, with lateral dimensions of \(\SI{2.04}{\micro\meter}\times\SI{2.04}{\micro\meter}\) and a design thickness of \(h=\SI{2.0}{\micro\meter}\), positioned above the photodetector plane.
(d) Scattering-operator description of the routing process. Each input wavelength \(\lambda_i\) is mapped to a distinct target region \(\Omega_i\) on the output monitor plane, where the incident-power-normalized output state \(e_i=e(\omega_i)\) satisfies \(\langle e_i, P_i e_i\rangle\ge\rho_0\). The target regions are mutually non-overlapping, \(\Omega_i\cap\Omega_j=\emptyset\) for \(i\neq j\).
}
\label{fig:concept}
\end{figure}

This photon-budget constraint becomes particularly acute when compact imagers
are extended beyond RGB acquisition toward multispectral and hyperspectral
sensing~\cite{hagen2013review,lapray2014multispectral,yuan2021snapshot}.
Many applications require wavelength-resolved measurements, including material
identification, biomedical sensing, precision agriculture, food inspection, and
environmental monitoring~\cite{goetz1985imaging,manolakis2002detection,lu2014medical,dale2013hyperspectral,gowen2007hyperspectral,adao2017hyperspectral}.
Conventional spectral imagers have enabled such capabilities using spatial or
spectral scanning, dispersive optics, tunable filters, and filter-array mosaics
combined with computational reconstruction~\cite{hagen2013review,sellar2005classification,gat2000imaging,lapray2014multispectral,zhou2024electrically}.
However, these architectures generally involve trade-offs among optical
throughput, acquisition speed, system volume, spatial sampling, and
reconstruction fidelity~\cite{hagen2013review,lapray2014multispectral,gehm2007single,wagadarikar2008single,yuan2021snapshot,monakhova2020spectral}.
Recent compact and integrated spectral imagers have pursued this balance through miniaturized optics and on-chip computational
sensing~\cite{mukhtar2025compact,bian2024broadband}.
For compact snapshot architectures, increasing the number of spectral channels within a fixed pixel size further tightens the trade-off among photon
utilization, spatial sampling, and spectral discrimination~\cite{jeon2019compact,lin2023metasurface,zhang2023handheld,catrysse2023spectral,mengu2023snapshot}.

For detector-integrated snapshot spectral imagers, spectral routers offer an alternative approach to addressing this trade-off~\cite{catrysse2023spectral,mengu2023snapshot}.
Rather than attenuating out-of-band photons, a spectral router uses a shared
nanophotonic structure to direct different wavelength components toward
designated detector sub-pixels~\cite{catrysse2023spectral,zou2022pixel,catrysse2022subwavelength}.
When implemented at the super-pixel level, such a device can perform wavelength
separation within a compact device size and can, in principle, improve photon
utilization relative to absorptive filter arrays~\cite{nishiwaki2013efficient,zou2022pixel,catrysse2022subwavelength,catrysse2023spectral}.
Adjoint-based topology optimization provides a design framework for
this task by optimizing a dielectric structure within a prescribed volume using
gradient-based full-wave simulations~\cite{lee2024inverse,kim2024freeform,
jeon2026inverse,molesky2018inverse,lalau2013adjoint,piggott2015inverse,
hammond2022high}.
Previous studies have demonstrated that filterless color splitters, color
routers, and related diffractive architectures can enhance photon
utilization~\cite{nishiwaki2013efficient,camayd2020multifunctional,miyata2021full,zhao2021perfect,zou2022pixel,li2022single,catrysse2022subwavelength,catrysse2023spectral}
and can extend wavelength-selective routing beyond RGB operation toward
snapshot multispectral imaging~\cite{catrysse2023spectral,mengu2023snapshot}.
Yet a basic question remains open: how does routing efficiency scale as
the number of simultaneously addressed spectral channels increases over a
fixed bandwidth?

Answering this question requires disentangling several effects that vary
together in a direct channel-count sweep. Increasing the number of channels over
a fixed spectral band simultaneously increases the number of routing
constraints, reduces the spacing between adjacent target wavelengths, and
modifies the local wavelength arrangement across neighboring sub-pixels.
This raises four questions. First, can a delay-bandwidth bound
perspective explain the efficiency degradation observed with increasing
channel count? Second, how does the minimum device thickness required to
maintain a prescribed worst-channel efficiency scale with the number of
channels, and conversely, what worst-channel efficiency can a finite-thickness
structure support as more channels are added? Third, how do the optimized
routing efficiencies of inverse-designed routers depend on the channel count
and the sub-pixel size? Fourth, is the degradation driven mainly by the number
of simultaneously routed channels, or by the accompanying changes in wavelength
spacing and local wavelength arrangement?

In this work, we address these questions by combining a delay-bandwidth
analysis with a systematic computational study of inverse-designed spectral
routers supporting 9, 16, 25, and 36 channels across the visible band. 
The bound analysis translates the routing condition into a minimum required
delay-bandwidth capacity that grows with both the number of channels and the
prescribed worst-channel efficiency. This capacity requirement is then expressed
as a minimum device-thickness estimate.
Adjoint-based topology optimization and finite-difference time-domain
(FDTD) simulations then quantify the routing efficiencies obtained at a fixed
device thickness and provide a numerical comparison with the trend indicated
by these estimates. Targeted control studies based on compressed wavelength spacing and
rearranged wavelength layouts further show that neither the reduced wavelength
spacing nor the local wavelength arrangement accounts for the observed
degradation, which instead tracks the number of simultaneously routed
channels. These results delineate a practical scaling
limit for spectral-router-based snapshot imaging.

\section{Problem Formulation and Delay-Bandwidth Bound}
\label{sec:problem_formulation}

To analyze wavelength routing in terms of delay-bandwidth capacity, we consider a finite-thickness spectral router based on a dielectric metasurface, in which a normally incident plane wave at each discrete wavelength channel is directed to a prescribed, spatially separated output region. The metasurface is modeled as a passive, linear, and time-invariant dielectric structure over the wavelength range of interest. The central question is how many spectrally distinct output states such a finite-thickness structure can support within a given bandwidth while maintaining a prescribed routing efficiency. To answer this question, we derive the minimum state-space separation imposed by the routing condition and compare the resulting trajectory-length requirement with delay-bandwidth limits for finite optical structures.

We define the routed spectrum in terms of $N$ discrete wavelength channels \(\lambda_i\) (\(i=1,\ldots,N\)), distributed over the spectral interval \([\lambda_{\min},\lambda_{\max}]\). The corresponding angular frequencies are given by \(\omega_i = 2\pi c_0/\lambda_i\), where \(c_0\) is the speed of light in vacuum. The incident-power-normalized output field state at \(\omega_i\) on the monitor plane is denoted by \(e_i=e(\omega_i)\), with \(\|e_i\|^2=T_{\mathrm{mon}}(\lambda_i)\le1\), where \(T_{\mathrm{mon}}=\mathcal{P}_{\mathrm{mon}}/\mathcal{P}_{\mathrm{inc}}\) is the ratio of total monitored output power to incident power. The target monitor regions \(\Omega_i\) are mutually non-overlapping, satisfying \(\Omega_i\cap\Omega_j=\emptyset\) for \(i\ne j\). In the analytical output-space model, \(P_i\) is the orthogonal projection operator onto \(\Omega_i\), with \(P_iP_j=0\) for \(i\ne j\), and \(\langle \cdot,\cdot\rangle\) denotes the positive-definite power inner product over the output field space. Since \(e_i\) is normalized to the incident power, \(\langle e_i,P_i e_i\rangle\) represents the fraction of the incident power delivered to the corresponding target region \(\Omega_i\). The routing condition is therefore written as
\begin{equation}
     \langle e_i,P_i e_i\rangle \ge \rho_0,
     \qquad i=1,\ldots,N,
     \label{eq:routing_condition_compact}
 \end{equation}
where \(\rho_0\) denotes the prescribed worst-channel incident-to-target routing efficiency threshold.

This routing condition imposes a minimum separation between output states assigned to different target regions. For \(i \neq j\), the state \(e_j\) contains at least a fraction \(\rho_0\) of the incident power in \(\Omega_j\), leaving at most a fraction \(1-\rho_0\) in \(\Omega_i\) because \(\|e_j\|^2\le1\). Applying the reverse triangle inequality separately to the components projected by \(P_i\) and \(P_j\) gives \(\|P_k(e_i-e^{i\theta}e_j)\|\ge\sqrt{\rho_0}-\sqrt{1-\rho_0}\) for \(k=i,j\) and any \(\theta\), provided \(\rho_0>1/2\). Orthogonality then gives \(\|e_i-e^{i\theta}e_j\|^2\ge2(\sqrt{\rho_0}-\sqrt{1-\rho_0})^2\). This gives the following bound on the phase-invariant distance between the two output states:
\begin{equation}
     d(e_i, e_j) \equiv \min_{\theta \in [0,2\pi)} \|e_i - e^{i\theta}e_j\| \ge \delta(\rho_0) = \sqrt{2 - 4\sqrt{\rho_0(1-\rho_0)}},
     \label{eq:distance_delta}
\end{equation}
where a nonzero separation, \(\delta(\rho_0) > 0\), requires \(\rho_0 > 1/2\). Otherwise, the same spatial pattern could in principle satisfy the routing condition for multiple disjoint target regions.

Although the routing condition is imposed only at discrete wavelength channels, the normalized output field state \(e(\omega)\) varies continuously with angular frequency \(\omega\). As \(\omega\) is swept across the band, the output state traces a continuous trajectory that passes through the \(N\) routed output states. The total length of this state-space trajectory is quantified by \(L_{\mathrm{actual}}\), defined as the integral of the phase-removed spectral derivative \(\|D_\omega e\|\). Here, \(D_\omega e\) denotes the phase-removed derivative, with \(\|D_\omega e\|=\min_{\alpha\in\mathbb R}\|de/d\omega-i\alpha e\|\). This removes only an arbitrary global phase variation, while changes in total monitored transmission remain included. Since the trajectory must connect \(N\) output states in spectral order, and each successive pair is separated by at least \(\delta(\rho_0)\), its length is bounded below by the sum of the corresponding chord distances:
\begin{equation}
     L_{\mathrm{actual}} = \int_{\omega_N}^{\omega_1} \|D_\omega e\| \, d\omega \ge (N-1)\delta(\rho_0).
     \label{eq:Lreq_compact}
 \end{equation}

To connect the required state-space trajectory length to the finite device thickness, we describe the metasurface as a frequency-dependent linear scattering system. The output state defined above can be written as \(e(\omega)=R_{\mathrm{mon}}S(\omega)u_0\) for the incident state \(u_0\), where \(S(\omega)\) is the full scattering operator, including all transmitted and reflected power-carrying channels, and \(R_{\mathrm{mon}}\) selects the monitored output. We use a fixed power-normalized channel representation with frequency-independent \(u_0\) and \(R_{\mathrm{mon}}\), \(\|u_0\|=1\), and \(\|R_{\mathrm{mon}}\|\le1\). For this step, the full scattering system is assumed lossless, so \(S\) is unitary. The spectral rate of change of this normalized output state can be bounded in terms of the Wigner-Smith time-delay operator, \(Q_{\mathrm{WS}} = -iS^\dagger dS/d\omega\), as \(\|D_\omega e\| \le \|Q_{\mathrm{WS}}\|\)~\cite{wigner1955lower, smith1960lifetime}. Here, operator norms are induced by the power inner product.

A common phase does not affect the trajectory length, so \(\|D_\omega e\|\le\|Q_{\mathrm{WS}}-\tau_{\mathrm{ref}}I\|\) for any real reference delay \(\tau_{\mathrm{ref}}(\omega)\). For a chosen reference delay, let \(\Delta T_{\max}\) be an upper bound on this relative-delay norm over the operating band. For the thickness estimate, we represent the relative-delay response by an effective longitudinal delay line of length \(h\), to which the selected delay-bandwidth bound is assumed to apply. We therefore bound the trajectory length achievable over the bandwidth as \(L_{\mathrm{actual}} \le \Delta T_{\max}\Delta\omega \le \kappa(h)\), where \(\kappa(h)\) denotes the dimensionless delay-bandwidth capacity of a structure with thickness \(h\)~\cite{presutti2020focusing}, and \(\Delta\omega=\omega_1-\omega_N\). Combining this upper bound with the lower-bound requirement in Eq.~\eqref{eq:Lreq_compact} yields the necessary condition within this delay-line model:
 \begin{equation}
     (N-1)\delta(\rho_0) \le \kappa(h).
     \label{eq:dbp_chain_compact}
 \end{equation}

To translate this condition into a thickness estimate, we next specify the delay-bandwidth capacity \(\kappa(h)\) for a finite-thickness metasurface. Because \(\kappa(h)\) depends on the physical model used to describe delay accumulation in the structure, we consider two representative bounds. First, we use a Tucker-type propagation-delay estimate for a waveguide-like, non-resonant dielectric structure~\cite{tucker2005slow}:
 \begin{equation}
     \kappa_{\mathrm{Tucker}}(h) = 2\pi\frac{h}{\lambda_c}\Delta n,
     \label{eq:tucker_capacity}
 \end{equation}
 where \(\lambda_c\) is the free-space center wavelength and \(\Delta n = n_{\max} - n_{\min}\) is the available refractive-index contrast. This form of the bound is appropriate when the spectral response can be interpreted as propagation through transparent dielectric segments with a well-defined group delay. Here, the constituent-index contrast bounds the modal effective-index range in the assumed waveguide model. The stated expression assumes a fractional bandwidth below unity.
 Substituting Eq.~\eqref{eq:tucker_capacity} into Eq.~\eqref{eq:dbp_chain_compact} yields the required thickness:
 \begin{equation}
     h_{\min}^{\mathrm{Tucker}} = \frac{\lambda_c}{2\pi\Delta n} (N-1)\delta(\rho_0).
     \label{eq:h_tucker_compact}
 \end{equation}

As a second estimate, we adopt Miller's delay-bandwidth bound, which applies to a broader class of one-dimensional delay structures without requiring transparent materials or a well-defined group velocity~\cite{miller2007fundamental}. In this formulation, the delay-bandwidth capacity is determined by the physical thickness of the structure and the relative permittivity contrast with respect to the background medium:
\begin{equation}
     \kappa_{\mathrm{Miller}}(h) = \frac{\pi}{\sqrt{3}}\frac{h}{\lambda_{c,b}}\eta_{\max},
     \label{eq:kappa_miller}
 \end{equation}
 where \(\lambda_{c,b}\) is the center wavelength measured in the background
medium, and \(\eta_{\max}=|\varepsilon_{\max}-\varepsilon_b|/\varepsilon_b\)
is the maximum relative permittivity contrast normalized by the background
permittivity \(\varepsilon_b\), following Miller's notation. The stated expression assumes a small fractional bandwidth and a device length large compared with \(\lambda_{c,b}\). Applying the
same necessary condition to Eq.~\eqref{eq:kappa_miller} gives the Miller-bound thickness estimate:
 \begin{equation}
     h_{\min}^{\mathrm{Miller}} = \frac{\sqrt{3}\lambda_{c,b}}{\pi\eta_{\max}} (N-1)\delta(\rho_0).
     \label{eq:h_miller_compact}
 \end{equation}

These estimates do not determine the relative contributions of propagation and resonant energy storage in the optimized structure.

\begin{figure*}[t!]
    \centering
    \includegraphics[width=1.0\linewidth]{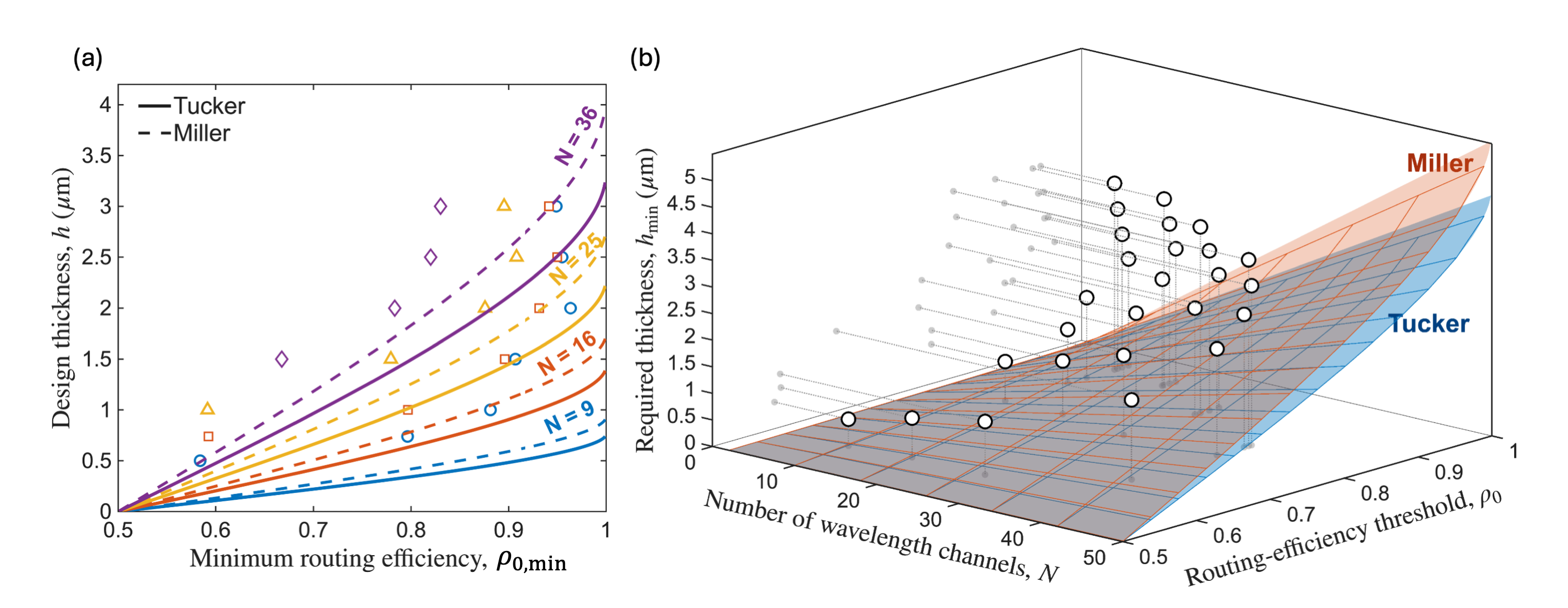}
    \caption{
    Comparison between simulated inverse-designed routers and Tucker-type~\cite{tucker2005slow} 
    and Miller-type~\cite{miller2007fundamental} delay-bandwidth thickness estimates.
    (a) Thickness estimates as functions of the routing-efficiency threshold for different channel counts \(N\). The markers denote the optimized spectral routers (this work) plotted using their achieved worst-channel efficiency \(\rho_{0,\min}\) and physical design thickness; at each thickness, the largest \(\rho_{0,\min}\) over the simulated sub-pixel sizes is shown (Supplementary Sec. S9).
    (b) Three-dimensional minimum-thickness estimates as functions of channel count \(N\) and prescribed threshold \(\rho_0\). The open markers indicate the optimized spectral router designs.
    }
    \label{fig:min_eff_vs_thickness}
\end{figure*}

\begin{table}[ht]
    \centering
    \caption{Delay-bandwidth thickness estimates for a TiO\(_2\)/SiO\(_2\) spectral router with \(N=36\) channels spanning \(400\)--\(700~\mathrm{nm}\). The Tucker-type estimate uses \(\lambda_c=509.09~\mathrm{nm}\) and \(\Delta n=1.20\), while the Miller-type estimate uses SiO\(_2\) as the background medium, with \(\lambda_{c,\mathrm{SiO_2}}=351.10~\mathrm{nm}\) and \(\eta_{\max}=2.3401\). The table reports \(\delta(\rho_0)\) and the corresponding minimum-thickness estimates for each prescribed threshold \(\rho_0\).
    }
    \label{tab:delay_bandwidth_bounds_compact}
    \begin{tabular}{c c c c}
        \hline
        \(\rho_0\) &
        \(\delta(\rho_0)\) &
        \(h_{\min}^{\mathrm{Tucker}}\) &
        \(h_{\min}^{\mathrm{Miller}}\) \\
        \hline
        0.8 & 0.6325 & \(1.49~\mu\mathrm{m}\) & \(1.83~\mu\mathrm{m}\) \\
        0.9 & 0.8944 & \(2.11~\mu\mathrm{m}\) & \(2.59~\mu\mathrm{m}\) \\
        1.0 & 1.4142 & \(3.34~\mu\mathrm{m}\) & \(4.09~\mu\mathrm{m}\) \\
        \hline
    \end{tabular}
\end{table}

We evaluate these thickness bounds for the TiO\(_2\)/SiO\(_2\) binary design region used in the optimized routers. Tucker's bound is evaluated using the TiO\(_2\)/SiO\(_2\) refractive-index contrast $(\Delta n = 1.20)$, whereas Miller's bound is evaluated by taking SiO\(_2\) as the background medium, giving \(\eta_{\max} = 2.3401\). For the \SIrange{400}{700}{\nano\meter} spectral range, the center wavelength is \(\lambda_c = 509.09~\mathrm{nm}\), corresponding to the wavelength at the midpoint of the frequency interval, and the background-medium center wavelength used for Miller's bound is \(\lambda_{c,b}=\lambda_{c,\mathrm{SiO_2}} = 351.10~\mathrm{nm}\). With these parameters, the resulting thickness estimates are summarized in Table~\ref{tab:delay_bandwidth_bounds_compact}.

Figure~\ref{fig:min_eff_vs_thickness} compares the simulated worst-channel routing efficiencies with the Tucker-type and Miller-type delay-bandwidth thickness estimates and visualizes their dependence on channel count and routing-efficiency threshold. In Fig.~\ref{fig:min_eff_vs_thickness}(a), the solid and dashed curves denote the theoretical Tucker and Miller estimates, respectively, while the markers indicate the simulated inverse-designed routers, plotted according to their achieved worst-channel efficiency \(\rho_{0,\min}\) and physical design thickness. Here, the achieved \(\rho_{0,\min}\) of each simulated design is used as the numerical counterpart to the prescribed threshold \(\rho_0\) in the state-space bound. The theoretical curves show that the minimum thickness required to support a prescribed routing-efficiency threshold increases with both the number of wavelength channels \(N\) and the target efficiency \(\rho_0\). This behavior follows directly from the required state-space trajectory length, \((N-1)\delta(\rho_0)\): increasing \(N\) introduces more routed output states that must be connected across the spectrum, while increasing \(\rho_0\) requires stronger separation between states assigned to different target regions.

The simulated designs are broadly consistent with this tendency: structures with larger \(N\) generally require greater thickness to attain comparable worst-channel efficiencies. However, the simulated efficiencies are not strictly monotonic with thickness. A larger design thickness increases the available optical degrees of freedom and can raise the attainable performance ceiling, but it does not guarantee that a particular optimization run will converge to a better design. Because the inverse-design problem is highly nonconvex, different thicknesses can lead to different local optima, and the optimized efficiency may vary with initialization, parameterization, and finite iteration count. Therefore, nonmonotonic variations among the simulated markers should not be interpreted as violations of the delay-bandwidth estimates. Rather, the bounds indicate the minimum capacity required to support a prescribed routing-efficiency threshold, whereas the markers correspond to specific numerically optimized designs obtained under finite computational constraints.

Figure~\ref{fig:min_eff_vs_thickness}(b) extends this comparison to the full \((N,\rho_0)\) parameter space by plotting the Tucker-type and Miller-type estimates as three-dimensional thickness-bound surfaces. Both surfaces increase with the number of wavelength channels \(N\) and the prescribed routing-efficiency threshold \(\rho_0\), showing that routing more channels or enforcing a higher worst-channel efficiency requires a larger delay-bandwidth capacity and therefore a greater device thickness. The simulated inverse-designed routers are overlaid using their achieved minimum efficiencies \(\rho_{0,\min}\) and physical design thicknesses. These markers are not expected to lie exactly on the theoretical surfaces, because the surfaces represent necessary delay-bandwidth capacity estimates rather than specific optimization outcomes. Nevertheless, the simulated results are consistent with the same overall tendency: routers with larger channel counts and higher achieved efficiencies lie in regions of larger thickness.

\section{Numerical Results}

This section presents the simulated routing performance of the optimized spectral routers. The delay-bandwidth analysis provides a qualitative explanation for the increasing difficulty of routing more spectral channels, while the efficiency trends are determined directly from controlled adjoint-based optimization and FDTD simulations. The numerical results are therefore interpreted independently of any quantitative fit to the theoretical estimates.

\subsection{Router Configurations and Evaluation Metrics}

We investigate inverse-designed spectral routers that direct each incident wavelength to a designated detector sub-pixel, enabling snapshot spectral imaging without mechanical scanning or absorptive filters. This approach builds on sensor-integrated wavelength-sorting devices, including color splitters and color routers, in which passive dielectric structures redistribute spectral components among detector regions rather than rejecting out-of-band light~\cite{nishiwaki2013efficient,zou2022pixel,
catrysse2022subwavelength,park2024towards,hong2024metasurface,choi2023optical}. Recent inverse-designed color routers and freeform metasurfaces have demonstrated that such wavelength-selective routing can be implemented within compact optical stacks compatible with image sensors~\cite{lee2024inverse,kim2024freeform,jeon2026inverse}. Here, we examine higher-channel-count spectral routers and quantify how their attainable routing efficiency varies with the number of simultaneously routed wavelengths.

We consider square super-pixel routers comprising \(N = k \times k\) sub-pixels, with \(N = 9\), 16, 25, and 36 corresponding to \(k = 3\), 4, 5, and 6, respectively. Each sub-pixel is assigned a target wavelength within the visible spectrum, and the investigated sub-pixel sizes cover submicrometer dimensions relevant to modern CMOS image sensors~\cite{fossum2016quanta,zou2022pixel}. As shown in Fig.~\ref{fig:concept}, the router consists of a finite-thickness binary TiO\(_2\)/SiO\(_2\) design region positioned above the detector plane. For each channel configuration, the dielectric distribution is optimized using adjoint-based topology optimization and evaluated by three-dimensional FDTD simulations~\cite{oskooi2010meep,hammond2022high}. Further details are provided in Supplementary Sec. S1.

We quantify the routing performance using the average routing efficiency,
\[
\bar{\eta}_{\mathrm{route}}
=
\frac{1}{N}
\sum_{i=1}^{N}
T_{ii}(\lambda_i),
\]
where \(T_{ii}(\lambda_i)\) is the fraction of incident power at wavelength \(\lambda_i\) delivered to its designated sub-pixel \(i\). Power directed to other sub-pixels is excluded, so this metric measures wavelength-selective routing rather than total transmission. The delay-bandwidth analysis uses a prescribed worst-channel efficiency threshold \(\rho_0\). For comparison with this criterion, the simulated worst-channel efficiency is defined as
\[
\rho_{0,\min}
=
\min_i T_{ii}(\lambda_i).
\]

For the channel-count scaling study, the sub-pixel size \(P\) is varied from \(\SI{0.12}{\micro\meter}\) to \(\SI{0.60}{\micro\meter}\). Unless stated otherwise, all routers have a thickness of \(h=\SI{2.0}{\micro\meter}\) and are simulated at an FDTD resolution of \(50~\text{pixels}/\si{\micro\meter}\). The wavelength-spacing and wavelength-arrangement control studies are performed at \(P=\SI{0.34}{\micro\meter}\), allowing direct comparison at a fixed sub-pixel size. For visualization, the focal-plane \(S_z\) distributions are cropped around their target sub-pixels and assembled into composite panels. Each map is normalized independently to its peak value solely for display. All routing efficiencies are calculated from the simulated power flux normalized to the incident power and are unaffected by this visualization procedure.

\begin{figure}[t]
\centering
\includegraphics[width=\linewidth]{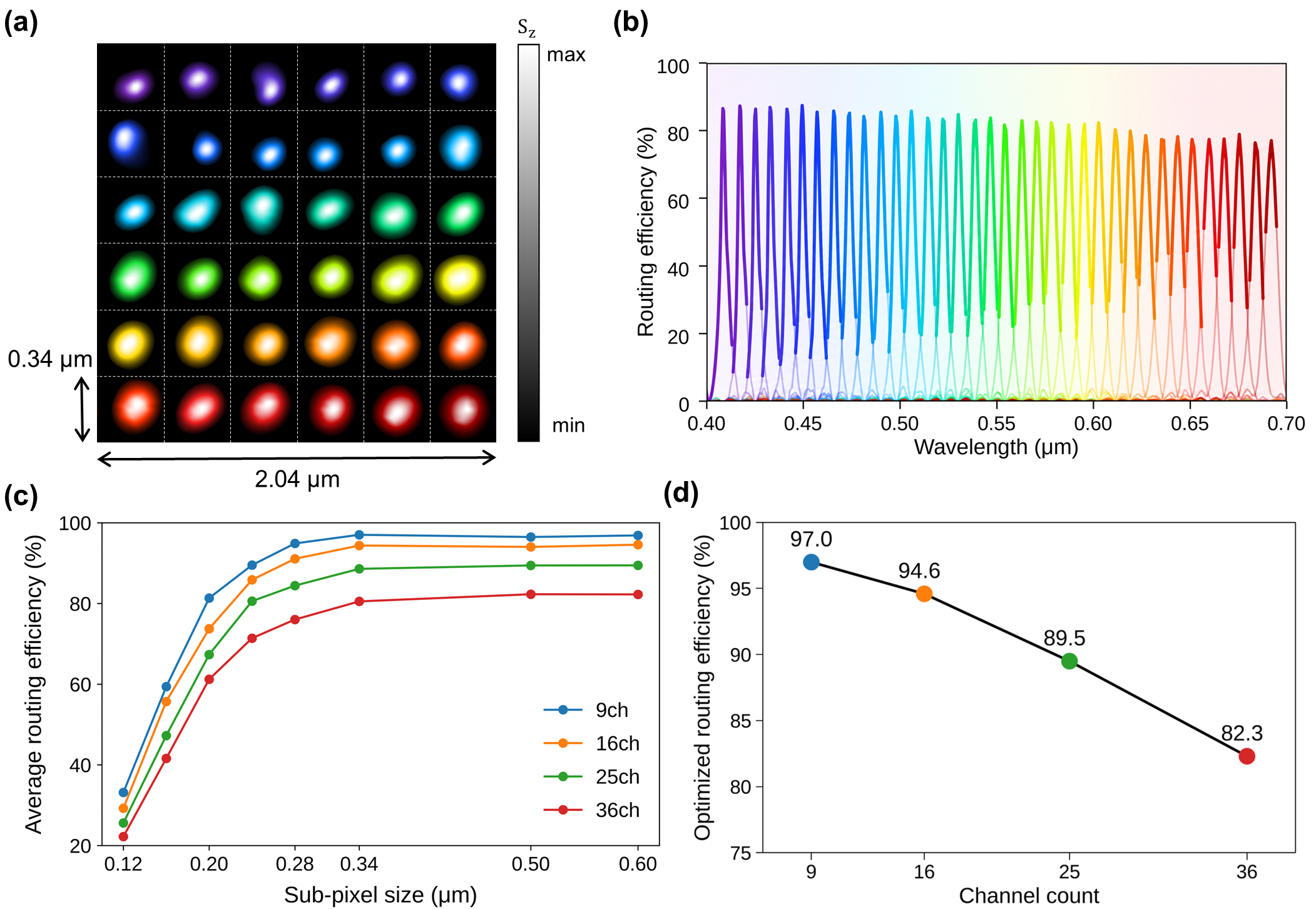}
\caption{
Channel-count-dependent scaling of routing efficiency.
(a) Composite focal-plane Poynting vector \(S_z\) distributions for the optimized 36-channel router at \(P=\SI{0.34}{\micro\meter}\), corresponding to a super-pixel size of \(\SI{2.04}{\micro\meter}\times\SI{2.04}{\micro\meter}\). Each channel map is cropped around its target sub-pixel and normalized independently to its peak value for visualization. 
(b) Routing-efficiency spectra of the 36-channel router across the visible band. Each curve corresponds to one target wavelength and its designated detector region.
(c) Average routing efficiency as a function of sub-pixel size \(P\) over \(\SIrange{0.12}{0.60}{\micro\meter}\) for 
\(N=9\), 16, 25, and 36 channels. For all channel counts, the efficiency increases with \(P\) and approaches a
plateau for \(P\ge\SI{0.34}{\micro\meter}\).
(d) Optimized routing efficiency, defined as the maximum average efficiency
among the near-plateau points
\(P\ge\SI{0.34}{\micro\meter}\), as a function of channel count. The efficiency decreases monotonically from \(97.0\%\) for
\(N=9\) to \(82.3\%\) for \(N=36\).
}
\label{fig:channel_count_scaling}
\end{figure}

\subsection{Efficiency Scaling with Sub-Pixel Size and Channel Count}

We evaluate the average routing efficiency for \(N=9\), 16, 25, and 36 spectral channels as the sub-pixel size \(P\) varies from \(\SI{0.12}{\micro\meter}\) to \(\SI{0.60}{\micro\meter}\). The resulting dependence on \(P\) and \(N\) is shown in Fig.~\ref{fig:channel_count_scaling}(c).

Snapshot spectral imaging beyond RGB has been demonstrated using several meta-optical approaches, including metasurface-enabled snapshot hyperspectral
imaging~\cite{xie2026snap}, six-channel spectral routers~\cite{catrysse2023spectral}, diffractive optical networks for 4, 9,
and 16 spectral bands~\cite{mengu2023snapshot}, and an eight-channel chromatic metalens array imager spanning the visible and near-infrared
range~\cite{audhkhasi2025single}. These studies establish high-channel-count operation but do not isolate the effect of channel count under matched design conditions. Here, the 9-, 16-, 25-, and 36-channel routers are optimized and evaluated using the same materials, device thickness, spectral bandwidth, and FDTD-based inverse-design procedure.

For all channel counts, the average routing efficiency increases with \(P\) and approaches a plateau for \(P\ge\SI{0.34}{\micro\meter}\), as the
larger lateral design area provides additional optical degrees of freedom. Two competing factors govern this behavior. Enlarging \(P\) increases both the
lateral extent of the design region and the separation between neighboring target sub-pixels, which relaxes the spatial confinement required of each
routed field. At the same time, the target regions must remain distinguishable on the scale of the operating wavelengths; when \(P\) falls well below this scale, diffraction prevents the routed power from being localized within a single target sub-pixel, and the efficiency drops rapidly. Beyond the plateau onset, further enlargement
yields only modest gains, indicating that the lateral dimension is no longer the limiting resource. Figure~\ref{fig:channel_count_scaling}(a,b) shows the
focal-plane \(S_z\) distributions and routing-efficiency spectra of the 36-channel router at \(P=\SI{0.34}{\micro\meter}\).

Within the near-plateau region, the average routing efficiency decreases monotonically with channel count, as shown in Fig.~\ref{fig:channel_count_scaling}(d). The maximum average efficiencies for \(P\ge\SI{0.34}{\micro\meter}\) are 97.0\%, 94.6\%, 89.5\%, and 82.3\% for \(N=9\), 16, 25, and 36, respectively. Under the matched design conditions considered here, this decline indicates that directing more wavelengths to distinct locations within a compact super-pixel imposes increasingly demanding routing constraints.

The achieved worst-channel efficiencies, \(\rho_{0,\min}\), are compared with the delay-bandwidth estimates in Section~\ref{sec:problem_formulation} and Fig.~\ref{fig:min_eff_vs_thickness}. For the fixed thickness \(h=\SI{2.0}{\micro\meter}\), the Miller estimate is more restrictive than the Tucker estimate for the TiO\(_2\)/SiO\(_2\) system. It permits unity worst-channel efficiency for \(N=9\) and 16, but limits the corresponding values to approximately 0.94 and 0.82 for \(N=25\) and 36, respectively. The simulated \(\rho_{0,\min}\) values remain below these estimates, with the higher-channel-count routers operating closer to the predicted capacity limit. This trend is consistent with the interpretation that increasing \(N\)
raises the delay-bandwidth demand of a finite-thickness router. The estimate, however, provides a necessary capacity condition rather than an exact performance bound for the three-dimensional optimized structures.

The following sections assess whether the observed efficiency degradation arises from reduced wavelength spacing, wavelength arrangement, or the increased number of simultaneously routed channels.

\begin{figure}[!t]
\centering
\includegraphics[width=\linewidth]{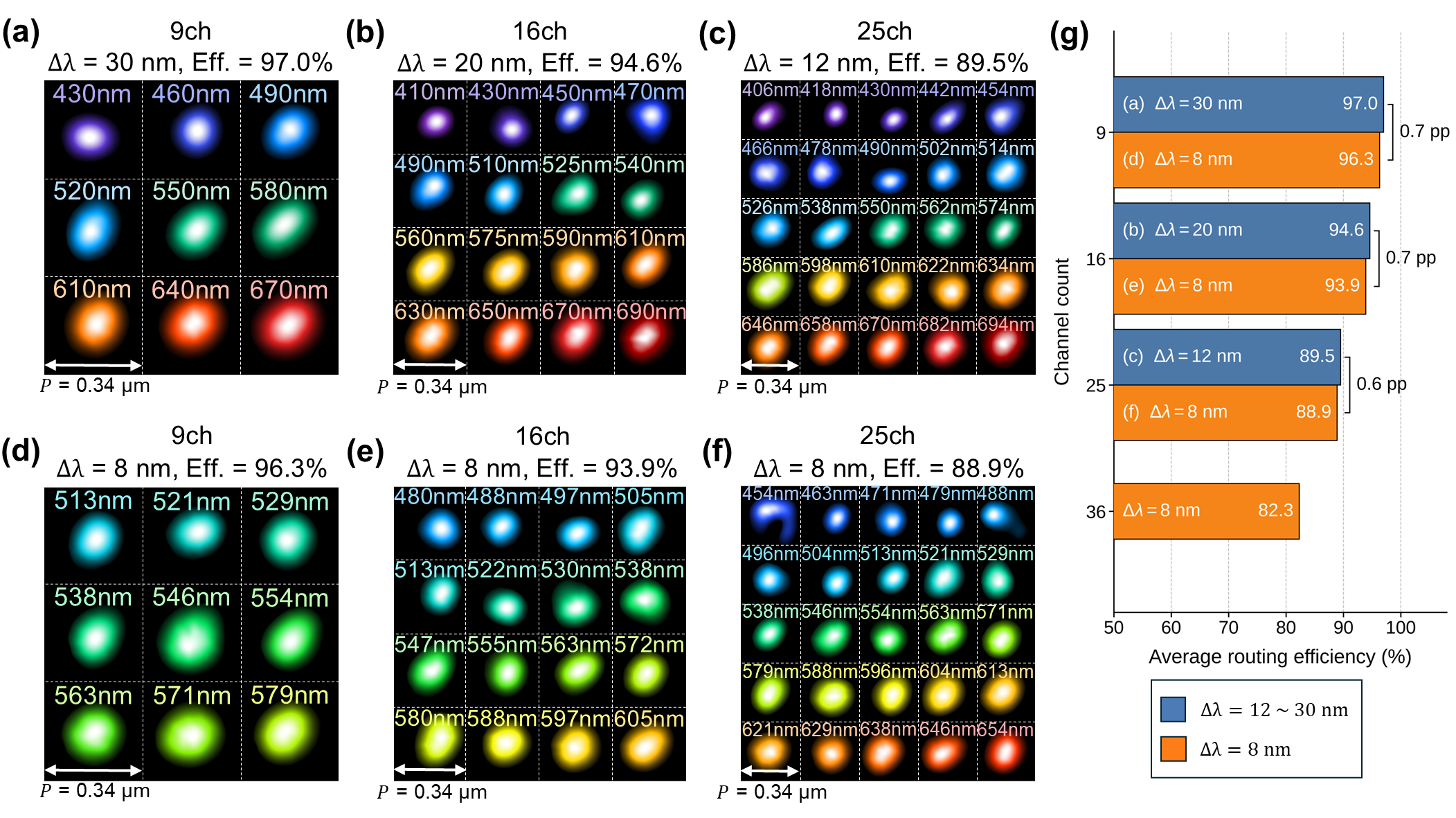}
\caption{
Effect of wavelength spacing on routing efficiency.
(a)--(c) Composite focal-plane \(S_z\) distributions for the 9-, 16-, and 25-channel routers using their channel-count-dependent wavelength spacings of \(\Delta\lambda=\SI{30}{\nano\meter}\), \(\SI{20}{\nano\meter}\), and \(\SI{12}{\nano\meter}\), respectively.
(d)--(f) Corresponding compressed-spacing designs with the same channel counts and spatial target layouts but a nominal spacing of \(\Delta\lambda=\SI{8}{\nano\meter}\).
Each channel map is cropped around its target sub-pixel and normalized independently to its peak value for visualization.
(g) Comparison of the average routing efficiencies for the original and compressed wavelength spacings. Reducing \(\Delta\lambda\) to \(\SI{8}{\nano\meter}\) lowers the efficiency by only \(0.6\)--\(0.7\) percentage points for the 9-, 16-, and 25-channel routers, substantially less than the decrease observed with increasing channel count. The 36-channel result is included as the \(\Delta\lambda=\SI{8}{\nano\meter}\) reference.}
\label{fig:channel_vs_spacing}
\end{figure}

\subsection{Channel Count versus Wavelength Spacing}

The channel-count sweep changes both the number of routing targets and the spacing between adjacent wavelengths. Across the \(\SI{400}{\nano\meter}\)--\(\SI{700}{\nano\meter}\) band, the baseline 9-, 16-, 25-, and 36-channel routers use nominal spacings of \(\Delta\lambda=\SI{30}{\nano\meter}\), \(\SI{20}{\nano\meter}\), \(\SI{12}{\nano\meter}\), and \(\SI{8}{\nano\meter}\), respectively. Integer-nanometer wavelength sampling introduces slight nonuniformity, with \(\SI{15}{\nano\meter}\) and \(\SI{20}{\nano\meter}\) steps in the 16-channel set and \(\SI{8}{\nano\meter}\) and \(\SI{9}{\nano\meter}\) steps in the 36-channel set. The efficiency decline observed as \(N\) increases may therefore arise from either the greater number of routing constraints or the narrower wavelength spacing.

A dependence on wavelength spacing is physically plausible because resolving closely spaced spectral channels generally requires a longer effective optical path or a narrower spectral response~\cite{yang2021miniaturization,su2018narrowband}. Dense spectral sampling is also relevant to snapshot spectral imaging; for example, a parallel metasystem resolves 20 channels over \(\SIrange{795}{980}{\nano\meter}\), corresponding to an average spacing of approximately \(\SI{10}{\nano\meter}\)~\cite{mcclung2020snapshot}. However, prior demonstrations vary channel spacing together with other device parameters and therefore do not isolate its effect at a fixed channel count.

To isolate the effect of wavelength spacing, we introduce the compressed-spacing controls shown in Fig.~\ref{fig:channel_vs_spacing}(d)--(f), with the corresponding full-super-pixel \(S_z\) maps provided in Supplementary Sec. S8. For each lower-channel-count router, the channel count and spatial target layout are unchanged, while the target wavelengths are reassigned with a nominal spacing of \(\SI{8}{\nano\meter}\), matching the 36-channel case and centered within the visible band. This comparison separates the influence of wavelength spacing from that of channel count. All compressed-spacing designs are evaluated at \(P=\SI{0.34}{\micro\meter}\), where the routing efficiency is near its plateau.

Compressing the wavelength spacing produces only a minor reduction in average routing efficiency. The matched-spacing references are the maximum near-plateau efficiencies reported in Fig.~\ref{fig:channel_count_scaling}(d), making this a conservative comparison with the compressed-spacing designs evaluated at \(P=\SI{0.34}{\micro\meter}\). The efficiency decreases from 97.0\% to 96.3\% for \(N=9\), from 94.6\% to 93.9\% for \(N=16\), and from 89.5\% to 88.9\% for \(N=25\), corresponding to reductions of 0.7, 0.7, and 0.6 percentage points, respectively. These changes are much smaller than the 14.7-percentage-point difference between the 9- and 36-channel routers. Thus, reducing the wavelength spacing to \(\SI{8}{\nano\meter}\) at fixed \(N\) does not account for the efficiency degradation observed with increasing channel count.

This result does not imply that wavelength spacing is unimportant in general. Within the spectral range and device conditions considered here, however, compressing the spacing to \(\SI{8}{\nano\meter}\) at fixed \(N\) produces a much smaller efficiency change than increasing the number of routed channels. The observed degradation therefore cannot be attributed to wavelength spacing alone.

\begin{figure}[!t]
\centering
\includegraphics[width=0.90\linewidth]{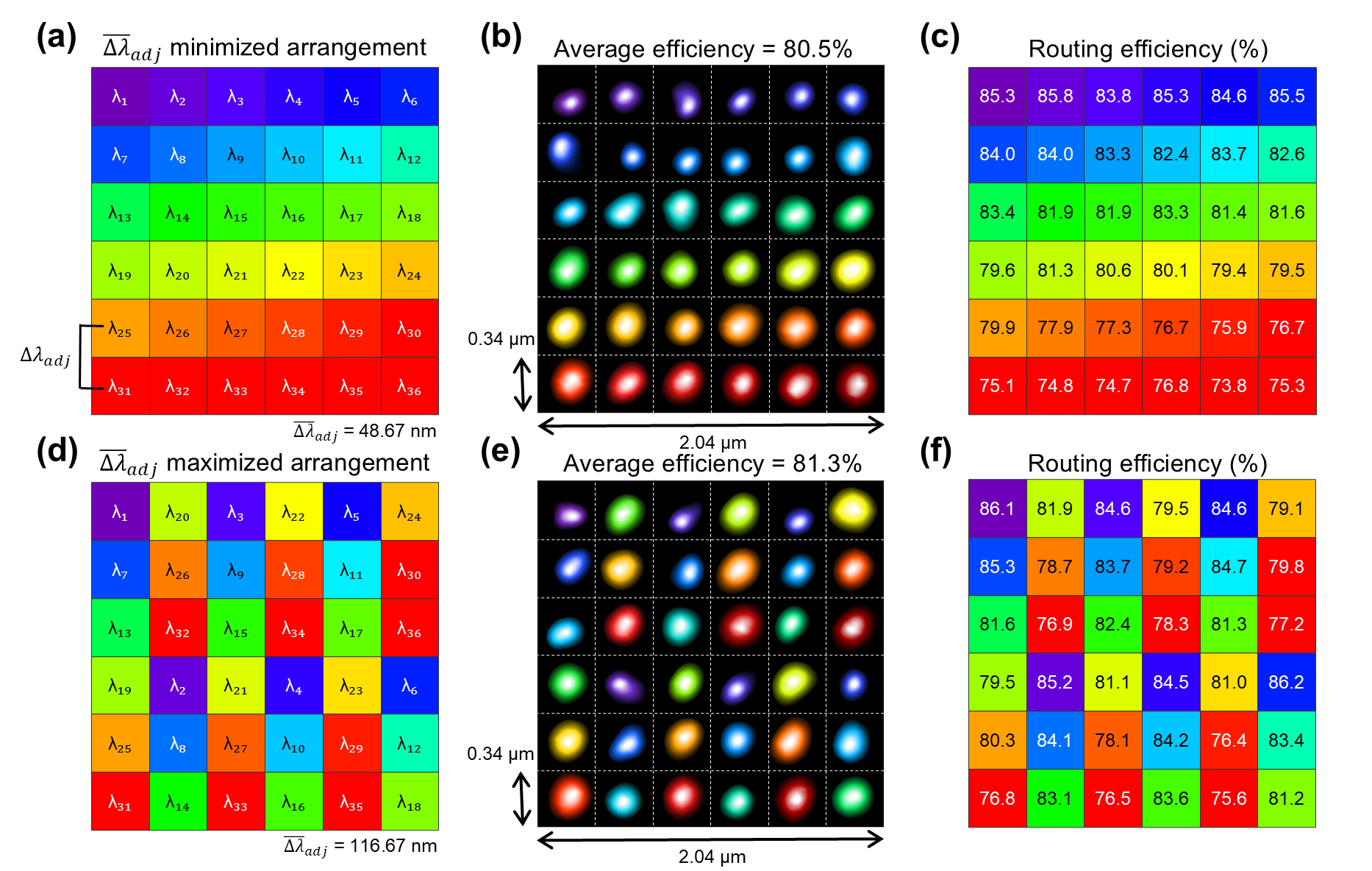}
\vspace{-0.4em}
\caption{
Limited effect of wavelength arrangement on routing efficiency. Two
36-channel arrangements are compared at \(P=\SI{0.34}{\micro\meter}\): the
\(\overline{\Delta\lambda}_{\mathrm{adj}}\)-minimized arrangement [(a)--(c)]
and the \(\overline{\Delta\lambda}_{\mathrm{adj}}\)-maximized arrangement
[(d)--(f)], where \(\overline{\Delta\lambda}_{\mathrm{adj}}\) denotes the
average wavelength difference between spatially adjacent sub-pixels.
(a),(d)~Wavelength-to-sub-pixel assignment maps, with
\(\overline{\Delta\lambda}_{\mathrm{adj}}=\SI{48.67}{\nano\meter}\) in (a)
and \(\SI{116.67}{\nano\meter}\) in (d).
(b),(e)~Corresponding composite focal-plane \(S_z\) distributions,
with average routing efficiencies of \SI{80.5}{\percent} and
\SI{81.3}{\percent}, respectively. Each channel map is cropped around its
target sub-pixel and normalized independently to its peak value for
visualization.
(c),(f)~Per-channel routing-efficiency maps arranged according to
the sub-pixel positions in (a) and (d). The average routing efficiency
changes by only 0.8 percentage points, indicating that wavelength
arrangement has only a minor effect compared with the
channel-count-dependent degradation observed in
Fig.~\ref{fig:channel_count_scaling}.
}
\label{fig:wavelength_arrangement}
\end{figure}

\subsection{Channel Count versus Wavelength Arrangement}

We next examine whether the spatial assignment of wavelengths affects routing efficiency. In the \(\overline{\Delta\lambda}_{\mathrm{adj}}\)-minimized arrangement, channels are assigned sequentially in wavelength, placing spectrally similar channels in neighboring sub-pixels. Such proximity could, in principle, increase susceptibility to local crosstalk between adjacent detector regions.

The influence of output-channel arrangement has also been examined in multichannel color routers. A recent diffractive-neural-network study tested ten randomly selected output configurations for a 9-channel router and reported little variation in average crosstalk or peak wavelength response~\cite{zhu2026new}. Here, we instead construct a deterministic arrangement that maximizes the mean spectral separation between adjacent sub-pixels and compare it directly with the sequential wavelength assignment.

To evaluate the effect of wavelength arrangement, we compare the \(\overline{\Delta\lambda}_{\mathrm{adj}}\)-minimized configuration of the 36-channel router with a configuration that maximizes the mean spectral separation between adjacent sub-pixels among the modular arrangements described in Supplementary Sec. S3. As shown in Fig.~\ref{fig:wavelength_arrangement}(a,d), \(\overline{\Delta\lambda}_{\mathrm{adj}}\) increases from \(\SI{48.67}{\nano\meter}\) to \(\SI{116.67}{\nano\meter}\). The corresponding focal-plane \(S_z\) distributions and per-channel routing efficiencies are presented in Fig.~\ref{fig:wavelength_arrangement}(b,e) and Fig.~\ref{fig:wavelength_arrangement}(c,f), respectively.

Despite the larger spectral separation between neighboring sub-pixels, the average routing efficiency changes only from \(\SI{80.5}{\percent}\) to \(\SI{81.3}{\percent}\), an increase of \(\num{0.8}\) percentage points. Thus, under the conditions tested, wavelength arrangement has only a minor influence on routing efficiency. This variation is substantially smaller than the efficiency reduction caused by increasing the number of simultaneously routed channels.

These results indicate that local wavelength adjacency is not the dominant limitation for the 36-channel router under the tested conditions. The larger performance penalty is instead associated with requiring a finite-thickness structure to map many spectrally distinct inputs to separate, non-overlapping detector regions.

\section{Discussion}

Among the variables examined at a fixed sub-pixel size, channel count had the strongest effect on routing efficiency. Compressing the wavelength spacing to \(\Delta\lambda=\SI{8}{\nano\meter}\) reduced the average efficiency by only \(0.6\)--\(0.7\) percentage points for the 9-, 16-, and 25-channel routers, while rearranging the 36 target wavelengths to increase the spectral separation between adjacent sub-pixels changed the efficiency by less than one percentage point. These variations are much smaller than the \(14.7\)-percentage-point decrease observed between the 9- and 36-channel designs. Under the material, thickness, bandwidth, and optimization conditions considered here, the efficiency degradation therefore cannot be attributed primarily to narrower wavelength spacing or local wavelength arrangement. Instead, it is associated mainly with the increasing number of wavelength-dependent field distributions that must be generated and directed to non-overlapping detector regions by a compact, finite-thickness dielectric structure.

This interpretation is consistent with the delay-bandwidth analysis. As the number of channels increases within a fixed spectral band, the router must generate more mutually distinct output states, increasing the required state-space trajectory length across frequency. A finite-thickness structure therefore needs greater delay-bandwidth capacity to maintain a given worst-channel efficiency. The observed decline in near-plateau efficiency is consistent with this increasing demand at fixed thickness. The bound should, however, be interpreted as a necessary capacity estimate that explains the trend rather than as a quantitative prediction of the optimized efficiencies.

These results have direct implications for compact snapshot spectral imagers. Increasing the channel count improves spectral sampling but reduces routing efficiency in compact super-pixel geometries. Enlarging the sub-pixel improves efficiency only until the response approaches a plateau for \(P\ge\SI{0.34}{\micro\meter}\). Further lateral scaling, therefore, provides diminishing returns. Practical designs should balance channel count, sub-pixel size, device thickness, and spectral bandwidth according to the required spectral resolution and available photon budget.

Several limitations should be considered. The reported routers were evaluated using idealized FDTD models that exclude fabrication errors, material variations, detector noise, and image-level spectral reconstruction. Supplementary Sec. S11 examines the effect of an imposed minimum feature size, but a full fabrication-tolerance analysis remains necessary. Because topology optimization may converge to local optima, the reported efficiencies represent attainable performance under matched computational conditions rather than global limits. The study is also restricted to normal incidence, leaving angular and polarization responses for future investigation. Supplementary Sec. S10 provides an initial extension to the visible--near-infrared range, while systematic studies of other spectral bands remain open.

Future studies should incorporate fabrication-aware optimization, experimental validation, and image-level reconstruction. Greater device thickness or designs with additional optical degrees of freedom may improve routing performance at high channel counts, while joint optimization of the router and reconstruction algorithm~\cite{choi2026meta,kang2026vlm} could better balance optical throughput, spectral discrimination, and reconstruction fidelity~\cite{liu2026high,seo2024deep,shen2023monocular}. These efforts are needed to determine the practical performance limits of spectral-router-based snapshot imaging systems.

\section{Conclusion}

We investigated how routing efficiency scales with channel count in inverse-designed spectral routers for compact snapshot spectral imaging. Using adjoint-based topology optimization and three-dimensional FDTD simulations, we designed TiO\(_2\)/SiO\(_2\) routers with 9, 16, 25, and 36 channels across the visible spectrum. The average routing efficiency increased with sub-pixel size and approached a plateau, while the maximum near-plateau efficiency decreased monotonically from 97.0\% for 9 channels to 82.3\% for 36 channels. Compressing the wavelength spacing to \(\Delta\lambda=\SI{8}{\nano\meter}\) or rearranging the wavelength-to-sub-pixel assignment changed the efficiency by less than one percentage point. Within the investigated design regime, the observed degradation is therefore associated primarily with the number of simultaneously imposed wavelength-routing constraints rather than with wavelength spacing or local wavelength arrangement. Together with the delay-bandwidth analysis, these results show that increasing spectral sampling in a compact, finite-thickness router results in a measurable loss of optical throughput. Practical designs must therefore balance channel count, sub-pixel size, device thickness, and operating bandwidth according to the spectral resolution and photon efficiency required by the imaging system.

\section*{Acknowledgments}
This work was supported by Samsung Electronics Co., Ltd. (IO260407-16222-01).
This work was also supported by the National Research Foundation of Korea (NRF)
grant funded by the Korean government (MSIT) (RS-2024-00338048,
RS-2025-25463760), the Culture, Sports and Tourism R\&D Program through the
Korea Creative Content Agency grant funded by the Ministry of Culture, Sports
and Tourism in 2024 (RS-2024-00332210), the Artificial Intelligence Graduate
School Program (No. RS-2020-II201373, Hanyang University) supervised by the
Institute of Information and Communications Technology Planning \& Evaluation
(IITP), and the artificial intelligence semiconductor support program to
nurture the best talents (IITP-(2025)-RS-2023-00253914) grant funded by the
Korean government (MSIT).

\bibliographystyle{unsrt}
\bibliography{references}

\end{document}